\documentclass[tightenlines,aps,prd,groupedaddress,showpacs,nofootinbib]
              {revtex4}%
\usepackage{graphicx}
\usepackage{amsmath}
\usepackage{bm}
\usepackage{relsize}
\RequirePackage{xspace}
\newcommand{\bqa}{\begin{eqnarray}}
\newcommand{\eqa}{\end{eqnarray}}

\begin{document}



\title{\mbox{}\\[10pt]
Radiative decays of bottomonia into charmonia and light mesons}


\author{Ying-Jia Gao$~^{(a)}$,Yu-Jie Zhang$~^{(a)}$, and Kuang-Ta
Chao$~^{(a)}$} \affiliation{ {\footnotesize (a)~Department of
Physics, Peking University,
 Beijing 100871, People's Republic of China}}




\begin{abstract}
In the framework of nonrelativistic QCD, we study the radiative
decays of bottomonia into charmonia, including $\Upsilon\to
\chi_{cJ}\gamma$, $\Upsilon\to \eta_c\gamma$, $\eta_b\to
J/\psi\gamma$, and $\chi_{bJ}\to J/\psi\gamma$. We give predictions
for their branching ratios with numerical calculations. E.g., we
predict the branching ratio for $\eta_b\to J/\psi\gamma$ is about
$1\times 10^{-7}$. As a phenomenological model study, we further
extend our calculation to the radiative decays of bottomonia into
light mesons by assuming the $f_2(1270)$, $f_2'(1525)$ and other
light mesons to be described by nonrelativistic $q\bar q ~(q=u,d,s)$
bound states with constituent quark masses. The calculated branching
ratios for $\Upsilon\to f_2(1270)\gamma$ and $\Upsilon\to
f_2'(1525)\gamma$ are roughly consistent with the CLEO data.
Comparisons with radiative decays of charmonium into light mesons
such as $J/\psi\to f_2(1270)\gamma$ are also given.  In all
calculations the QED contributions are taken into account and found
to be significant in some processes.

\end{abstract}

\pacs{12.38.Bx; 13.25.Hw; 14.40.Gx}

\maketitle


\section{Introduction}

Radiative decays of bottomonium (e.g.~$\Upsilon, ~\eta_b,
~\chi_{bJ}$) into charmonium are expected to be described by
nonrelativistic quantum chromodynamics (NRQCD), since both
bottomonium and charmonium  are made of heavy quark and heavy
antiquark, and are nonrelativistic bound states. For heavy
quarkonium decay and production, the rates can be factorized into a
short-distance part, which can be calculated in QCD perturbatively,
and a long-distance part, which are governed by nonperturbative QCD
dynamics~\cite{BBL}. Therefore, radiative decays of bottomonium into
charmonium may provide a useful test for NRQCD factorization, which
is assumed to hold also for these specific exclusive processes, and
may also provide some practical estimates for decays such as
$\eta_b\to J/\psi\gamma$, which might be useful in search for the
not yet discovered $\eta_b$ meson. As a phenomenological model
study, we further extend our calculation to the radiative decays of
bottomonia into light mesons by assuming the $f_2(1270)$,
$f_2'(1525)$ and other light mesons to be described by
nonrelativistic $q\bar q ~(q=u,d,s)$ bound states with constituent
quark masses $m_q` (q=u,~d)= 350~MeV, ~m_s=500~MeV$ as in
constituent quark models. These radiative decays are known as the
gluon rich channels, and regarded as a good place to investigate the
interactions between quarks and gluons in these OZI forbidden
processes, and there have been some earlier work discussing these
processes (see, e.g.,\cite{kor,Krammer:1978qp}). In this paper, as
our previous work~\cite{CPL060725}, we will perform a complete
numerical calculation for the quark-gluon loop diagrams involved in
these processes, and we will also include contributions from QED
diagrams in the same processes.

We adopt the assumption that both heavy quarkonium and light mesons
are described by the color-singlet non-relativistic wave functions.
Based on this assumption, we study $\Upsilon\to \chi_{cJ}\gamma$,
$\Upsilon\to \eta_c\gamma$, $\Upsilon\to f_J\gamma$, $\Upsilon\to
\eta\gamma$, $J/\psi\to f_J\gamma$, $J/\psi\to \eta\gamma$,
$\chi_{bJ}\to J/\psi(\rho,\omega,\phi)\gamma$ and $\eta_b\to
J/\psi(\rho,\omega,\phi)\gamma$ etc.

The rest of this paper is as follows. In section II, we will give
the descriptions and  main techniques in our calculations, and then
make predictions for the decay rates of $\Upsilon\to
\chi_{cJ}\gamma$, $\Upsilon\to \eta_c\gamma$, $\eta_b\to
J/\psi\gamma$, and $\chi_{bJ}\to J/\psi\gamma$. Then, in the
following section, we will generalize this method to those processes
in which the final states are light mesons. Finally, we will summary
all the results in section IV.

\section{Bottomonium radiative decays to charmonium}
In this section, we will study the radiative decays of bottomonium
into charmonium. In NRQCD, heavy quarkonium wave function is
described by a Fock state expansion in terms of the relative
velocity $v$ between the quark and antiquark, and the leading term
is a color-singlet $Q\bar Q$ state , which has the same quantum
numbers as the physical heavy quarkonium. In certain processes, the
non-leading terms with color-octet $Q\bar Q$ pair and soft gluons
may make dominant contributions. E.g., in the $\Upsilon$ radiative
decays to light quark jets $\Upsilon\to q\bar q\gamma$ the
color-octet contribution could be larger that the color-singlet
contribution (depending on the estimates of the color-octet matrix
elements)~\cite{CTP06}. In the radiative decays of bottomonium into
charmonium, the short distance transitions of a color-octet $b\bar
b$ into a color-octet $c\bar c$ by emitting a photon are shown in
Fig.1, where $q=c$, and $q\bar q$ are in color-octet $(^3S_1)^8$ or
$(^3P_J)^8$. Compared with the case of $\Upsilon$ radiative decays
to light quark jets $\Upsilon\to q\bar q\gamma$ (see
Ref.\cite{CTP06} for an estimate of the color-octet contributions),
here the contribution of color-octet $c\bar c$ is greatly suppressed
by the smallness of the color-octet matrix elements of $(^3S_1)^8$
or $(^3P_J)^8$ (note that the color-octet matrix element of
$(^3S_1)^8$ is only 1\% of that of color-singlet $(^3S_1)^1$ for
$J/\psi$. Therefore, we will neglect the color-octet contributions,
and only concentrate on the color-singlet description of heavy
quarkonia in the following calculations.

\begin{figure*}[t]
\includegraphics[width=11cm]{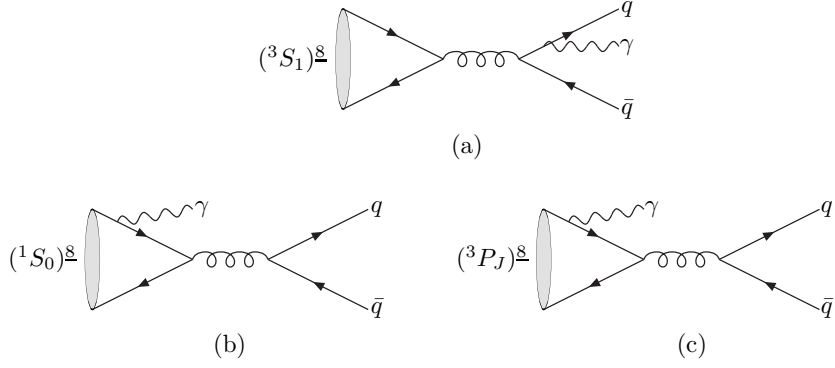}
\caption{Feynman diagrams for transitions from color octet $b\bar b$
to color-octet $q\bar q$ (where $q=c$) by emitting a photon}
\label{fvs}
\end{figure*}

\subsection{General results}
In the nonrelativistic approximation, the $\Upsilon$ radiative decay
into a color-singlet $c \bar c$ pair, which subsequently hadronizes
into charmonium, can be described by the diagrams in Fig.2, and the
amplitude can be expressed as\cite{liu}

\begin{eqnarray}
\label{amp2}   &&\hspace{-2.0cm}{\cal
A}\Big(b\bar{b}({}^{3}S_{1})(2p_b)\rightarrow c\bar{c}({}^{2S
+1}L_{J})(2p_c)\Big)=
\sqrt{C_{L_\Upsilon}}\sqrt{C_L}\sum\limits_{L_{\Upsilon z}
S_{\Upsilon z} }\sum\limits_{s_1s_2 }\sum\limits_{jk}
\sum\limits_{L_z S_z }\sum\limits_{s_3
s_4}\sum\limits_{il}\nonumber\\
&\times& \langle 1 \mid \bar{3}k;3j \rangle \langle J_\Upsilon
J_{\Upsilon z} \mid L_\Upsilon L_{\Upsilon z };S_\Upsilon
S_{\Upsilon z} \rangle \langle S_\Upsilon S_{\Upsilon z} \mid
s_1;s_2 \rangle \nonumber\\
&\times&\langle s_3;s_4\mid S S_z\rangle\langle L L_z ;S S_z\mid
J J_z\rangle\langle 3l;\bar{3}i\mid 1\rangle\nonumber\\
 &\times&\left\{
\begin{array}{ll}
{\cal A}\Big(b_j(p_b)+\bar{b}_k(p_b)\rightarrow
 \gamma(p_3)+ c_l(p_c)+\bar{c}_i(p_c)\Big)&(L=S),\\
\epsilon^*_{\alpha}(L_Z) {\cal
A}^\alpha\Big(b_j(p_b)+\bar{b}_k(p_b)\rightarrow
 \gamma(p_3)+ c_l(p_c)+\bar{c}_i(p_c)\Big)
&(L=P),
\end{array}
\right.
\end{eqnarray}
where $\langle 3l;\bar{3}i\mid 1\rangle$$=\delta_{li}/\sqrt{N_c}$~,~
$\langle s_1;s_2\mid S S_{z}\rangle$~, and $ \langle L L_z ;S
S_z\mid J J_z\rangle$ are respectively the color-SU(3), spin-SU(2),
and angular momentum Clebsch-Gordan coefficients for $Q\bar{Q}$
pairs projecting on appropriate bound states. ${\cal
A}(b_j(p_b)+\bar{b}_k(p_b)\rightarrow Q_l(p_c)+\bar{Q}_i(p_c))$ is
the decay amplitude for $Q\bar{Q}$ production and ${\cal A}^\alpha$
is the derivative of the amplitude with respect to the relative
momentum between the quark and anti-quark in the bound state. The
coefficients $C_{L_\Upsilon}$ and $C_L$ can be related to the radial
wave function of the bound states or its derivative with respect to
the relative spacing as
\begin{equation}
\label{cs} C_S=\frac{1}{4\pi}|R_s (0)|^2,\ \ \
C_P=\frac{3}{4\pi}|R_p'(0)|^2.
\end{equation}

 \begin{figure*}[t]
\includegraphics[width=16cm]{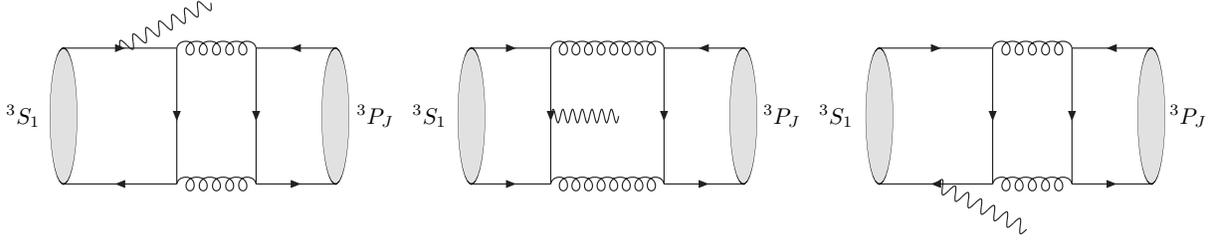}
\caption{Typical QCD Feynman diagrams for heavy quarkonium ($^3S_1$)
radiative decays into ($^3P_J$) mesons} \label{fvs}
\end{figure*}

The spin projection operators $P_{SS_z}(p,q)$ which describe
production of quarkonium are expressed in terms of quark and
anti-quark spinors as\cite{liu,Kuhn}:
 \bqa P_{SS_z}(p,q)\!=\!\!\!\sum_{s_1,s_2}\!\!v(\frac{p}{2}\!-\!q,\!s_2)
 \bar u(\!\frac{p}{2}\!+\!q,\!s_1) \!\langle s_1;\!s_2|SS_z\!\rangle,
 \eqa
We list the spin projection operators and their derivatives with
respect to the relative momentum, which will be used in the
calculations, as
\begin{eqnarray}
P_{00}(p,0)&=&\frac{1}{2\sqrt{2}}\gamma_{5}(\not{\! p}+2m),\\
P_{1S_Z}(p,0)&=&\frac{1}{2\sqrt{2}}\not{\epsilon}^*(S_z)(\not{\! p}+2m),\\
P_{1S_z}^{\alpha}(p,0)&=&\frac{1}{4\sqrt{2}m}
[\gamma^{\alpha}\not{\epsilon}^*(S_z)(\not{\! p}+2m)- (\not{\!
p}-2m)\not{\epsilon}^*(S_z)\gamma^{\alpha}].
\end{eqnarray}
And the spin projection operators which describe the annihilation of
quarkonium are the complex conjugate of the corresponding operators
for production. The polarization vectors for the ${}^3P_J$ states
are shown below:

 \bqa\label{pol}
\!\!\!\!\sum_{L_Z
S_Z}\!\!\varepsilon^{\ast\alpha}\!(\!L_z\!)\epsilon^{\ast\beta}\!(\!S_z\!)
\langle 1L_z;\!1S_z|1J_z\rangle\!&=&\! \frac{-i
\epsilon^{\alpha\beta\lambda\kappa}p_\kappa\epsilon_\lambda^\ast\!(\!J_z\!)}
{\sqrt{2}M},\\
\!\!\!\!\sum_{L_Z
S_Z}\!\!\varepsilon^{\ast\alpha}\!(\!L_z\!)\epsilon^{\ast\beta}\!(\!S_z\!)
\langle
1L_z;\!1S_z|0\,0\,\rangle\!&=&\!\!\frac{1}{\sqrt{3}}(-\!g^{\alpha\beta}\!+\!\frac{p^\alpha
p^\beta}{M^2}\!), \\
\!\!\!\!\sum_{L_Z
S_Z}\!\!\varepsilon^{\ast\alpha}\!(\!L_z\!)\epsilon^{\ast\beta}\!(\!S_z\!)
\langle
1L_z;\!1S_z|2J_z\rangle\!&=&\!\epsilon^{\ast\alpha\beta}(J_z),
 \eqa
where p is the momentum of P-wave quarkonium, and M is the mass of
the corresponding quarkonium. $\epsilon_\lambda(J_z)$ are the
polarization vectors for $J=1$. $\epsilon^{\alpha\beta}(J_z)$ are
the polarization vectors for $J=2$, which are symmetric under the
exchange $\alpha\leftrightarrow\!\beta$.

\begin{figure}
\includegraphics[width=7cm,height=3cm]{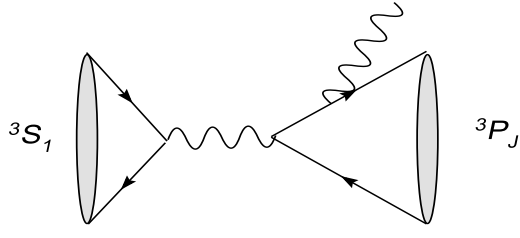}
\caption{Typical QED Feynman diagrams for heavy quarkonium ($^3S_1$)
radiative decays into ($^3P_J$) mesons} \label{fvsQED}
\end{figure}

The QCD Feynman diagrams of $\Upsilon \to \gamma \eta_c(\chi_{cJ})$,
in which the $c\bar c$ are produced through gluons, are shown in
FIG.\ref{fvs}, while the QED Feynman diagrams, in which the $c\bar
c$ are produced through the photon, are shown in FIG.\ref{fvsQED}.
In the calculation, we use {\tt FeynCalc} \cite{FeynCalc} for the
tensor reduction and {\tt LoopTools}\cite{LoopTools} for the
numerical evaluation of infrared safe integrals. We follow the way
in Ref.\cite{fourrank} to deal with five-point functions and high
tensor loop integrals that can not be calculated by {\tt LoopTools}
and {\tt FeynCalc}, such as
\bqa
 \!\!\!\!\!E_{\alpha\beta\rho\sigma}&=&\int\mathrm{{d}}^Dk
\frac{k_\alpha k_\beta k_\rho k_\sigma}{k^2[(k+p_c)^2-m_c^2](k+2
p_c)^2 [(p_b+k)^2-m_b^2][(k+2 p_c-p_b)^2-m_b^2]}.
\eqa
where $p_c$
is the momentum of $c$ quark, $p_b$  the momentum of $b$ quark,
$m_c$ the charm quark mass, and $m_b$ the bottom quark mass.


\begin{table}[tb]
\begin {center}
\begin{tabular}{|c|c|c|c|c|}
 \hline
process&$\Upsilon\to\chi_{c2}\gamma$&$\Upsilon\to\chi_{c1}\gamma$
&$\Upsilon\to\chi_{c0}\gamma$&$\Upsilon\to\eta_c\gamma$\\\hline
$BR_{QCD}$&$5.1\times 10^{-6}$&$4.5\times 10^{-6}$&$4.0\times 10^{-6}$&$2.9\times 10^{-5}$\\
$BR_{QCD+QED}$&$5.6\times 10^{-6}$&$9.8\times 10^{-6}$&$3.2\times
10^{-6}$&$4.9\times 10^{-5}$\\
\hline\hline process&$\chi_{b2}\to J/\psi\gamma$&$\chi_{b1}\to
J/\psi\gamma$&$\chi_{b0}\to J/\psi\gamma$&$\eta_b\to
J/\psi\gamma$\\\hline
$\Gamma_{QCD}$(GeV)&$2.7\times 10^{-10}$&$3.8\times 10^{-10}$&$5.0\times 10^{-10}$&$2.8\times 10^{-9}$\\
$\Gamma_{QCD+QED}$(GeV)&$3.6\times 10^{-10}$&$3.7\times
10^{-10}$&$1.3\times 10^{-10}$&$9.6\times 10^{-10}$\\
$\Gamma_{QED}$(GeV)&$3.8\times 10^{-11}$&$3.3\times
10^{-12}$&$1.3\times 10^{-10}$&$1.2\times 10^{-9}$\\
\hline
\end{tabular}
\caption{Decay widths and branching ratios for radiative decays of
bottomonium into charmonium. The decay widths $\Gamma$ are in units
of GeV, and the branching ratios $BR$ are given for the $\Upsilon$.}
 \label{table1}
\end {center}
\vspace{-0.5cm}
\end{table}

In the numerical calculations, the quark masses are taken to be
$m_b=4.7\,\mbox{GeV}$, $m_c=1.5\,\mbox{GeV}$, the wave functions at
the origin can be found from potential model calculations in
Ref.\cite{Quig}: $|\mathcal{R}^{b \bar b}_S(0)|^2=6.477 ~$GeV$^3$,
$|\mathcal{R}^{c \bar c}_S(0)|^2=0.81 ~$GeV$^3$,
$|\mathcal{R}^{\prime b \bar b}_P(0)|^2=1.417 ~$GeV$^5$,
${|\mathcal{R}^{\prime c \bar c}}_P(0)|^2=0.075 ~$GeV$^5$. In the
bottomonium decay, the strong coupling constant is chosen as
$\alpha_s(2m_b)=0.19$. The numerical results of radiative decays of
bottomonium into charmonium are listed in Table
\uppercase\expandafter{\romannumeral 1}.

\subsection{$\eta_b$ radiative decay to $J/\psi$}

The $\eta_b$ meson is the only one among the low lying bottomonium
states that has not been observed experimentally. To search for the
$\eta_b$ meson a number of decay channels have been suggested, e.g.,
decays into the $J/\psi J/\psi$ and $D\overline
D^{(*)}$~\cite{braaten,maltoni,jia}. In any case, the radiative
decay $\eta_b\to J/\psi\gamma$ should be a useful channel for the
$\eta_b$ in view of the cleanness of the signal (this possibility
has also been considered in Ref.\cite{hao}).

From Table \uppercase\expandafter{\romannumeral 1}, we can see that
for the $\eta_b$ decay $\eta_b\to J/\psi\gamma$ the QCD and QED
contributions are comparable but destructive, and, as a result, the
decay width of $\eta_b\to J/\psi\gamma$ is only $9.6\times
10^{-10}$~GeV. In order to know the branching ratio of this decay
channel, we should have an estimate for the $\eta_b$ total width. In
fact, we can estimate its total width through
$\Gamma_{tot}(\eta_b)\approx \Gamma(\eta_b \to gg)$ \cite{BBL}. For
$\Gamma(\eta_b \to gg)$, with next to leading order (NLO) QCD
radiative corrections, we have
\begin{equation}
\label{eq:etabgg} \Gamma(\eta_b \to gg)={|\mathcal{R}_s(0)|^2 C_F
\alpha_s^2(2m_b)\over 2m_b^2}
 \Bigg\{ 1 + \left[ \left( {\pi^2 \over 4} - 5
\right) C_F
        + \left( {199 \over 18} -  {13 \pi^2 \over 24} \right) C_A
        - {8 \over 9} n_f \right] {\alpha_s \over \pi} \Bigg\}.
\end{equation}
With the parameters used above, we can get
$\Gamma_{tot}(\eta_b)\approx 11.4$~MeV. Then the branching ratio is
$Br(\eta_b \to \gamma J/\psi) \approx 8.4\times 10^{-8}$. If we use
the leading order formula in Eq.(\ref{eq:etabgg}), the decay width
is $\Gamma_{tot}(\eta_b)\approx 7.1~$MeV, and the branching ratio
becomes $Br(\eta_b \to \gamma J/\psi) \approx 1.4\times 10^{-7}$.

On the other hand, with the spin symmetry in the nonrelativistic
limit ($v=0$), the $\eta_b$ wave function at the origin
$|\mathcal{R}_s(0)|^2$ can be determined from the $\Upsilon$
leptonic width,
\begin{eqnarray}
\label{eq:Upsee}
 \Gamma(\Upsilon\to e^{+}e^{-})&=&N_c
Q_b^2 \alpha^2 \frac{|\mathcal{R}_s(0)|^2}{3 m_b^2}\left(1-\frac{16
\alpha_s}{3 \pi}\right),
\end{eqnarray}
and the $\eta_b$ total width is then related to the $\Upsilon$
leptonic width,
\begin{eqnarray}
\Gamma_{tot}(\eta_b)=\frac{3 C_F
\alpha_s^2(2m_b)}{2N_cQ_b^2\alpha^2}\frac{ 1 + \left[ \left( {\pi^2
\over 4} - 5 \right) C_F
        + \left( {199 \over 18} -  {13 \pi^2 \over 24} \right) C_A
        - {8 \over 9} n_f \right] {\alpha_s \over \pi} }{1-\frac{16 \alpha_s}{3
\pi}}\Gamma(\Upsilon\to e^{+}e^{-}).
\end{eqnarray}
Using $m_b=4.7$~GeV, $\alpha_s(2m_b)=0.19$, and experimental data
$\Gamma(\Upsilon\to e^{+}e^{-})=1.340\pm0.018$~KeV\cite{Pdg}, we can
get $\Gamma_{tot}(\eta_b)\approx 13.0$~MeV.  Then the branching
ratio is $Br(\eta_b \to \gamma J/\psi) \approx 7.3\times 10^{-8}$.
If we use the leading order formula in Eq.(\ref{eq:etabgg}) and
Eq.(\ref{eq:Upsee}), the $\Gamma_{tot}(\eta_b)\approx 5.45~$MeV, the
branching ratio is $Br(\eta_b \to \gamma J/\psi) \approx 1.7\times
10^{-7}$.

In any case, we find that the branching ratio $Br(\eta_b \to \gamma
J/\psi)$ is of order $1\times 10^{-7}$. This small number makes it
quite difficult to search for $\eta_b$ through this decay channel.

\subsection{Helicity ratios with $\chi_{c1}$ and $\chi_{c2}$}

We give predictions for branching ratios for different helicity
states in $\Upsilon\to\chi_{cJ}\gamma$ decays. As in Ref.
\cite{JPMa}, we choose the moving direction of $\chi_{cJ}$ as the
z-axis, and introduce three polarization vectors:
 \bqa
\omega^{\mu}(1)&=&\frac{-1}{\sqrt{2}}(0,1,i,0),\,\,\,\nonumber\\
\omega^{\mu}(-1)\!\!&=&\!\!\frac{1}{\sqrt{2}}(0,1,-i,0)\nonumber\\
\omega^{\mu}(0)&=&\frac{1}{m}(|\textbf{k}|,0,0,k^0), \eqa
thus we can characterize the tensor $\epsilon^{\mu \nu}(\lambda)$ of
$\chi_{c2}$
 \bqa
\!\!\!\!\!\!\!\epsilon^{\alpha\beta}(2)\!\!\!&=&\!\!\!\omega^{\alpha}(1)\omega^{\beta}(1)\nonumber\\
\!\!\!\!\!\!\!\epsilon^{\alpha\beta}(1)\!\!\!&=&\!\!\!\frac{1}{\!\!\sqrt{2}}(\omega^{\alpha}(1)\omega^{\beta}(0)
+\omega^{\alpha}(0)\omega^{\beta}(1))\nonumber\\
\!\!\!\!\!\!\epsilon^{\alpha\beta}(0)\!\!\!&=&\!\!\!\frac{1}{\!\!\sqrt{6}}
(\omega^{\alpha}(-1)\omega^{\beta}(1)+2
\omega^{\alpha}(0)\omega^{\beta}(0)
+\omega^{\alpha}(1)\omega^{\beta}(-1))\nonumber\\
\epsilon^{\alpha\beta}(-1)\!\!\!&=&\!\!\!\frac{1}{\!\!\sqrt{2}}(\omega^{\alpha}(0)\omega^{\beta}(-1)
+\omega^{\alpha}(-1)\omega^{\beta}(0))\nonumber\\
\epsilon^{\alpha\beta}(-2)\!\!\!&=&\!\!\!\omega^{\alpha}(-1)\omega^{\beta}(-1)
\eqa
The helicity ratios are introduced as
\begin{equation*}
\begin{split}
x^2 = 
\frac{|a_1|^2}{|a_0|^2} \text{ and } 
y^2 = 
\frac{|a_2|^2}{|a_0|^2},
\end{split}
\end{equation*}
where $a_{\lambda},\ \lambda=0,\ 1,\ 2,$ are the normalized helicity
amplitudes, which satisfy $|a_0|^2+|a_1|^2+|a_2|^2 = 1$.  Namely,
$|a_{\lambda}|^2$ is the probability of the final state meson with
helicity $\pm\lambda$. Then the ratios $x^2$ and $y^2$ only depend
on the mass ratio $m_c/m_b$. With the same choice of parameters, we
predict ratios for different helicities in Table
\uppercase\expandafter{\romannumeral 2}.

\begin{table}[tb]
\begin {center}
\begin{tabular}{|c|c|c|c|c|c|}
 \hline
&QCD&QCD+QED&&QCD&QCD+QED
\\\hline
$x^2(\Upsilon\to\gamma \chi_{c2})$&0.37&0.38
&$x^2(\Upsilon\to\gamma \chi_{c1})$&0.064&0.075\\
$y^2(\Upsilon\to\gamma \chi_{c2})$&0.14&0.14&&&\\
           \hline
\end{tabular}
\caption{Results for $\Upsilon\to\chi_{cJ}\gamma (J=1,2)$ with
different helicity states}
 \label{table2}
\end {center}
\vspace{-0.5cm}
\end{table}

\section{Radiative decays of heavy quarkonium into light mesons}
As a purely phenomenological model-dependent study, In this section
we will extend our calculations performed above for radiative decays
of bottomonia into charmonia to the radiative decays of bottomonia
into light mesons. Our assumption is that the light mesons such as
the $f_2(1270)$, $f_2'(1525)$, and $f_1(1285)$  can be described by
nonrelativistic $q\bar q ~(q=u,d,s)$ bound states with constituent
quark masses.

 In the numerical calculations, the light quark masses are taken to
 be
 $m_s=0.50\,\mbox{GeV}$, $m_u=m_d=0.35\,\mbox{GeV}$.
The parameters for the heavy quarks are the same as that
used in section \uppercase\expandafter{\romannumeral 2}
,$|\mathcal{R}^{b \bar b}_S(0)|^2=6.477 ~$GeV$^3$, $|\mathcal{R}^{c
\bar c}_S(0)|^2=0.81 ~$GeV$^3$, $|\mathcal{R}^{\prime b \bar
b}_P(0)|^2=1.417 ~$GeV$^5$, $|\mathcal{R}^{ \prime c \bar
c}_P(0)|^2=0.075 ~$GeV$^5$, $m_b=4.7~$GeV, and $m_c=1.5~$GeV. The
strong coupling constant is chosen as $\alpha_s=0.19$ and
$\alpha_s=0.26$ in bottomonium and charmonium decays respectively.

As widely accepted assignments we assume that $f_2(1270)$ and
$f_1(1285)$ are mainly composed of $(u\bar u+d\bar d)/\sqrt{2}$
(neglecting the mixing with $s\bar s$ for simplicity). But for
$f_0(980)$, there are many possible assignments such as the
tetraquark state, the $K\bar K$ molecule, and the P-wave $s\bar s$
dominated state (for related discussions on $f_0(980)$ and other
scalar mesons, see, e.g., the topical review--note on scalar mesons
in \cite{Pdg} and \cite{close}). Since experimental data show that
$D_s^{+}\to f_0(980)\pi^{+}$ has a large branching ratio
(BR)\cite{Pdg}, here we assign $f_0(980)$ as an $s\bar s$ dominated
P-wave state as a tentative choice (we do not try to justify this
assignment).

As to the wave functions at the origin of light mesons, it is very
difficult to determine them without any doubt. Using the theoretical
expression for the widths of $f_2\to\gamma\gamma$\cite{BBL},

 \begin{eqnarray}
 \label{eq:f21270}
\Gamma_{f_2(1270)\to \gamma\gamma}^{(th)}&=&\frac{6 N_c}{5}
(Q_u^2+Q_d^2)^2 \alpha^2 \frac{|\mathcal{R}'_P(0)|^2}{m^4}(1-\frac{8
\alpha_s}{3 \pi})^2 \nonumber \\
 \Gamma_{f_2'(1525)\to
\gamma\gamma}^{(th)}&=&\frac{12 N_c}{5} Q_s^4 \alpha^2
\frac{|\mathcal{R}'_P(0)|^2}{m^4}(1-\frac{8 \alpha_s}{3 \pi})^2,
\end{eqnarray}

where $N_c$=3 is the color number, $\alpha=1/137$, and fitting them
with their experimental values 2.6~KeV and 0.081~KeV for $f_2(1270)$
and $f_2'(1525)$ respectively\cite{Pdg}, we get
\begin{eqnarray}
|\mathcal{R}^{\prime n \bar n}_P(0)|^2& =&1.58\times 10^{-3}\,\, \mbox{GeV}^5, \nonumber \\
  |\mathcal{R}^{\prime s \bar s}_P(0)|^2
&=&2.23\times 10^{-3}\,\, \mbox{GeV}^5.
\end{eqnarray}
If we use the leading order formula in Eq(\ref{eq:f21270}), then
\begin{eqnarray}
|\mathcal{R}^{\prime n \bar n}_P(0)|^2& =&6.6\times 10^{-4}\,\,
\mbox{GeV} ^5\nonumber \\
 |\mathcal{R}^{\prime s \bar
s}_P(0)|^2& =&1.1\times 10^{-3}\,\, \mbox{GeV}^5.
\end{eqnarray}

For the vector mesons, the wave functions at the origin may be
determined from their leptonic decay $V\to e^{+}e^{-}$~
($V=\phi,\rho$) widths. Using
\begin{eqnarray}\label{eq:rhophi} \Gamma(\phi(1020)\to e^{+}e^{-})&=&N_c
Q_s^2 \alpha^2 \frac{|\mathcal{R}(0)|^2}{3 m_s^2}(1-\frac{8
\alpha_s}{3 \pi})^2
= (1.27\pm 0.04)~\mbox{KeV},
\end{eqnarray}
we can get
\begin{eqnarray}
|\mathcal{R}^{ n \bar n}_S(0)|^2& =&0.11\,\,
\mbox{GeV}^3,\hspace{1cm} \nonumber \\ |\mathcal{R}^{ s \bar
s}_S(0)|^2 &=&0.19\,\, \mbox{GeV}^3.
\end{eqnarray}
If we use the leading order formula in Eq.(\ref{eq:rhophi}), then
\begin{eqnarray}
|\mathcal{R}^{ n \bar n}_S(0)|^2& =&0.032\,\,
\mbox{GeV}^3,\hspace{1cm} \nonumber \\ |\mathcal{R}^{ s \bar
s}_S(0)|^2 &=&0.054\,\, \mbox{GeV}^3.
\end{eqnarray}

The wave functions at the origin of light mesons can also be
determined from potential models \cite{Quigg:1979vr}. From
experimental data, $\Delta E=M(2S)-M(1S)$ is $675~$MeV, $638~$MeV,
$661~$MeV, $589~$MeV, and $563~$MeV for $\rho$, $\omega$, $\phi$,
$J/\psi$, and $\Upsilon$ respectively. In the logarithmic potential,
$\Delta E$ is independent of quark masses. So we may select the
logarithmic potential, which gives
\begin{eqnarray}
|R_S(0)|^2 &\propto& m_q^{3/2}\nonumber \\
|R_P(0)|^2 &\propto& m_q^{5/2}
\end{eqnarray}
With $|\mathcal{R}^{c \bar c}_S(0)|^2=0.81 ~$GeV$^3$ and
$|\mathcal{R}^{ \prime c \bar c}_P(0)|^2=0.075 ~$GeV$^5$
\footnote{The wavefunction at the origin in the logarithmic
potential for $c \bar c$ is $|\mathcal{R}^{c \bar c}_S(0)|^2=0.815
~$GeV$^3$ and $|\mathcal{R}^{ \prime c \bar c}_P(0)|^2=0.078
~$GeV$^5$. It is consistent with the B-T potential result that was
used here.}. Then we can get
\begin{eqnarray}
|\mathcal{R}^{n \bar
n}_S(0)|^2&=&\left(\frac{m_n}{m_c}\right)^{3/2}|\mathcal{R}^{ c \bar
c}_S(0)|^2=0.091\mbox{GeV}^3 \nonumber \\
|\mathcal{R}^{s \bar
s}_S(0)|^2&=&\left(\frac{m_s}{m_c}\right)^{3/2}|\mathcal{R}^{ c \bar
c}_S(0)|^2=0.156\mbox{GeV}^3 \nonumber \\
|\mathcal{R}^{\prime n \bar
n}_P(0)|^2&=&\left(\frac{m_n}{m_c}\right)^{5/2}|\mathcal{R}^{\prime
c \bar
c}_P(0)|^2=1.97 \times 10^{-3}\mbox{GeV}^5 \nonumber \\
|\mathcal{R}^{\prime s \bar
s}_P(0)|^2&=&\left(\frac{m_s}{m_c}\right)^{5/2}|\mathcal{R}^{\prime
c \bar c}_P(0)|^2=4.81 \times 10^{-3}\mbox{GeV}^5\end{eqnarray}

In the numerical calculation, the parameters are taken to be
$|\mathcal{R}^{s \bar s}_S(0)|^2=0.054 ~$GeV$^3$, $|\mathcal{R}^{n
\bar n}_S(0)|^2=0.032 ~$GeV$^3$, $|\mathcal{R}^{\prime s \bar
s}_P(0)|^2=1.1\times 10^{-3} ~$GeV$^5$, $|\mathcal{R}^{\prime n \bar
n}_P(0)|^2=6.6\times 10^{-4} ~$GeV$^5$. The numerical results are
shown in Table \uppercase\expandafter{\romannumeral 3} and  Table
\uppercase\expandafter{\romannumeral 4}.


\begin{table}[tb]
\begin {center}
\begin{tabular}{ |c|c|c|c|c|c|c|}
 \hline
process &$\Upsilon\to\gamma f_0(n \bar n)$&$\Upsilon\to\gamma
f_1(1285)$&$\Upsilon\to\gamma f_2(1270)$&$\Upsilon\to\gamma
f_0(980)$&$\Upsilon\to\gamma f_1'(1420)$&$\Upsilon\to\gamma
f_2'(1525)$
\\\hline
$BR_{th}^{QCD}$&$7.2\times 10^{-5}$&$2.6\times 10^{-5}$&$7.1\times
10^{-5}$
&$2.2\times 10^{-5}$&$1.0\times 10^{-5}$&$2.2\times 10^{-5}$\\
$BR_{th}^{QCD+QED}$&$6.4\times 10^{-5}$&$3.9\times 10^{-5}$
&$6.3\times 10^{-5}$&$2.0\times 10^{-5}$&$1.2\times 10^{-5}$&$2.0\times 10^{-5}$\\
$BR_{ex}$&$\backslash$&$\backslash$&$1.00\pm0.10 \times 10^{-4}$
&$<3\times 10^{-5}$&$\backslash$&$3.7^{+1.2}_{-1.1}\times 10^{-5}$\\
 \hline
process &$J/\psi\to\gamma f_0(n \bar n)$&$J/\psi\to\gamma
f_1(1285)$&$J/\psi\to\gamma f_2(1270)$&$J/\psi\to\gamma
f_0(980)$&$J/\psi\to\gamma f_1'(1420)$&$J/\psi\to\gamma f_2'(1525)$
\\\hline
$BR_{th}^{QCD}$&$2.0\times 10^{-3}$&$2.0\times 10^{-3}$&$2.5\times
10^{-3}$
&$6.7\times 10^{-4}$&$7.5\times 10^{-4}$&$8.5\times 10^{-4}$\\
$BR_{th}^{QCD+QED}$&$1.8\times 10^{-3}$&$2.7\times 10^{-3}$
&$2.4\times 10^{-3}$&$6.5\times 10^{-4}$&$8.5\times 10^{-4}$&$8.5\times 10^{-4}$\\
$BR_{ex}$&$\backslash$&$(6.1\pm0.8)\times10^{-4}$&$(13.8\pm1.4)\times10^{-4}$&$\backslash$
&$(7.9\pm1.3)\times10^{-4}$&$(4.5^{+0.7}_{-0.4})\times10^{-4}$\\
\hline
\end{tabular}
\caption{Numerical results for $\Upsilon(J/\psi)\to \gamma f_J$ .}
 \label{table1}
\end {center}
\vspace{-0.5cm}
\end{table}

\begin{table}[tb]
\begin {center}
\begin{tabular}{ |c|c|c|c|c|}
 \hline
process&$\chi_{b2}\to\gamma\rho$&$\chi_{b1}\to\gamma\rho$&$\chi_{b0}\to\gamma\rho$&$\eta_b\to\gamma\rho$\\\hline
 $\Gamma_{QCD}$(GeV)&$1.1\times 10^{-10}$&$2.2\times 10^{-10}$&$3.5\times 10^{-11}$&$1.2\times 10^{-10}$\\
 $\Gamma_{QCD+QED}$(GeV)&$6.8\times 10^{-10}$&$2.2\times 10^{-10}$&$8.4\times 10^{-10}$&$1.1\times 10^{-8}$\\ \hline
process&$\chi_{b2}\to\gamma\omega$&$\chi_{b1}\to\gamma\omega$&$\chi_{b0}\to\gamma\omega$&$\eta_b\to\gamma\omega$\\\hline
 $\Gamma_{QCD}$(GeV)&$1.2\times 10^{-11}$&$2.4\times 10^{-11}$&$3.8\times 10^{-12}$&$1.3\times 10^{-11}$\\
 $\Gamma_{QCD+QED}$(GeV)&$7.6\times 10^{-11}$&$2.4\times 10^{-11}$&$9.3\times 10^{-11}$&$1.2\times 10^{-9}$\\ \hline
process&$\chi_{b2}\to\gamma\phi$&$\chi_{b1}\to\gamma\phi$&$\chi_{b0}\to\gamma\phi$&$\eta_b\to\gamma\phi$\\\hline
 $\Gamma_{QCD}$(GeV)&$3.6\times 10^{-11}$&$5.8\times 10^{-11}$&$1.6\times 10^{-11}$&$5.9\times 10^{-11}$\\
  $\Gamma_{QCD+QED}$(GeV)&$1.3\times 10^{-10}$&$5.8\times 10^{-11}$&$7.5\times 10^{-11}$&$1.2\times 10^{-9}$\\\hline
process&$\chi_{c2}\to\gamma\rho$&$\chi_{c1}\to\gamma\rho$&$\chi_{c0}\to\gamma\rho$&$\eta_c\to\gamma\rho$\\\hline
$BR_{th}^{QCD}$&$1.3\times 10^{-5}$&$4.1\times 10^{-5}$&$3.2\times
10^{-6}$ &$1.5\times 10^{-6}$\\
$BR_{th}^{QCD+QED}$&$3.8\times 10^{-5}$&$4.2\times
10^{-5}$&$2.0\times 10^{-6}$ &$4.5\times 10^{-6}$\\\hline
process&$\chi_{c2}\to\gamma\omega$&$\chi_{c1}\to\gamma\omega$&$\chi_{c0}\to\gamma\omega$&$\eta_c\to\gamma\omega$\\\hline
$BR_{th}^{QCD}$&$1.5\times 10^{-6}$&$4.6\times 10^{-6}$&$3.5\times
10^{-7}$ &$1.7\times 10^{-7}$\\
$BR_{th}^{QCD+QED}$&$4.2\times 10^{-6}$&$4.7\times
10^{-6}$&$2.2\times 10^{-7}$ &$5.0\times 10^{-7}$\\\hline
process&$\chi_{c2}\to\gamma\phi$&$\chi_{c1}\to\gamma\phi$&$\chi_{c0}\to\gamma\phi$&$\eta_c\to\gamma\phi$\\
\hline $BR_{th}^{QCD}$&$3.3\times 10^{-6}$&$1.1\times
10^{-5}$&$1.3\times
10^{-6}$ &$7.1\times 10^{-7}$\\
$BR_{th}^{QCD+QED}$&$6.5\times 10^{-6}$&$1.1\times
10^{-5}$&$3.0\times 10^{-8}$ &$4.0\times
10^{-7}$\\\hline\end{tabular} \caption{Predicted decay widths for
$\chi_{bJ}(\eta_b)\to\gamma \rho(\omega,\phi)$.}
 \label{table1}
\end {center}
\vspace{-0.5cm}
\end{table}

The branching ratio of $\Upsilon$ radiative decay into a light meson
is smaller than the corresponding branching ratio of $J/\psi$ by a
factor of
\begin{eqnarray}
\frac{BR(\Upsilon \to \gamma\  M)}{BR(J/\psi \to \gamma\  M)} \sim
\left(\frac{Q_b}{Q_c}\right)^2\left(\frac{m_c}{m_b}\right)^2\frac{\alpha(2m_b)}{\alpha(2m_c)}
\sim 0.02
\end{eqnarray}
We can find this theoretical ratio is $0.013 \sim 0.036$. The
experimental ratio is 0.072 for $f_2(1270)$, and 0.082 for
$f_2'(1525)$. The corresponding theoretical ratios are $0.026$ and
$0.024$ for $f_2(1270)$ and  $f_2'(1525)$ respectively. If we use
larger constituent quark masses, e.g. $m_u=m_d=M(1270)/2$,
$m_s=M(1525)/2$, the ratios are $0.018$ and $0.018$ for  $f_2(1270)$
and  $f_2'(1525)$ respectively.

With the same parameters, we give the branching ratios for different
helicity states in Table  \uppercase\expandafter{\romannumeral 5}.
The corresponding values of helicity parameters $x$ and $y$ are
$x=0.46$, $y=0.23$. Recently, new experimental data for the
contributions of different helicities in process $J/\psi\to\gamma
f_2(1270)$ have been given by the BES Collaboration \cite{besjpsi}:
$x=0.89 \pm 0.02 \pm 0.10$ and $y=0.46 \pm 0.02 \pm 0.17$ (see
also\cite{Pdg}). It is about 2 times larger than our results. But if
we use a larger constituent quark mass, e.g. $m_u=M(1270)/2$, we
will get substantially increased values $x=0.79$ and $y=0.58$ (also
see Ref.\cite{kor}). We emphasize that the helicity parameters are
very sensitive to the light quark masses, and hence very useful in
clarifying the decay mechanisms. Note that if $m_u/m_c\to 0$, we
will have $x\to 0$ and $y\to 0$, but this is inconsistent with data.

%

\begin{table}[tb]
\begin {center}
\begin{tabular}{|c|c|c|c|c|c|c|c|c|c|}
 \hline
$m_q$(GeV)&&QCD&QCD+QED&&QCD&QCD+QED&&QCD&QCD+QED\\\hline
0.35&$x^2(\Upsilon\to \gamma f_2)$&0.023&0.023 &$y^2(\Upsilon\to
\gamma f_2)$&0.0013&0.0014&$x^2(\Upsilon\to\gamma
f_1)$&0.00069&0.00010\\           \hline 0.635&$x^2(\Upsilon\to
\gamma f_2)$&0.073&0.075 &$y^2(\Upsilon\to \gamma
f_2)$&0.0095&0.010&$ x^2(\Upsilon\to \gamma f_1)$&0.0047&0.0067\\
\hline 0.50&$x^2(\Upsilon\to\gamma f_2')$&0.045&0.046
&$y^2(\Upsilon\to\gamma f_2')$&0.0043&0.0045&$ x^2(\Upsilon\to\gamma
f_1')$&0.0018&0.0023\\           \hline 0.763&$x^2(\Upsilon\to\gamma
f_2')$&0.10&0.11 &$y^2(\Upsilon\to\gamma f_2')$&0.017&0.018&$
x^2(\Upsilon\to\gamma f_1')$&0.0087&0.010\\           \hline
0.35&$x^2(J/\psi\to \gamma f_2)$&0.21&0.21 &$y^2(J/\psi\to \gamma
f_2)$&0.055&0.057&$x^2(J/\psi\to\gamma f_1)$&0.026&0.029\\
\hline 0.635&$x^2(J/\psi\to \gamma f_2)$&0.62&0.62 &$y^2(J/\psi\to
\gamma f_2)$&0.33&0.33&$ x^2(J/\psi\to \gamma f_1)$&0.13&0.14\\
\hline 0.50&$x^2(J/\psi\to\gamma f_2')$&0.40&0.40
&$y^2(J/\psi\to\gamma f_2')$&0.16&0.16&$ x^2(J/\psi\to\gamma
f_1')$&0.072&0.074\\           \hline 0.763&$x^2(J/\psi\to\gamma
f_2')$&0.86&0.86 &$y^2(J/\psi\to\gamma f_2')$&0.57&0.57&$
x^2(J/\psi\to\gamma f_1')$&0.21&0.22\\           \hline
\end{tabular}
\caption{Results for $\Upsilon\to\chi_{cJ}\gamma$ with different
helicity states}
 \label{table2}
\end {center}
\vspace{-0.5cm}
\end{table}

\section{Summary}
In this paper, we mainly investigate the radiative decays of
bottomonium into charmonium, such as $\Upsilon\to \chi_{cJ}\gamma$,
$\chi_{bJ}\to J/\psi\gamma$, $\eta_b\to J/\psi\gamma$ and
$\Upsilon\to \eta_c\gamma$ based on the NRQCD approach. Based on our
numerical calculations, we predict that the branching ratios of
$\Upsilon\to \chi_{cJ}\gamma$ and decay widths of $\chi_{bJ}\to
J/\psi\gamma$.
All the above processes are perturbative calculable, and it is a
good way to test NRQCD.

We next focus on the cases of heavy quarkonium radiative decays into
light mesons, including $\Upsilon(J/\psi)\to f_J\gamma$ and
$\chi_{bJ}\to \rho(\omega,\phi)\gamma$ et ac.

In this work, we also find that the QED effects in some radiative
processes are really significant. For $\Upsilon\to \gamma \chi_{cJ}$
decay, the pure electromagnetic process only affects the final
results for $J=0,2$ a little, but for the $J=1$ state the result may
change by a factor of 2. The same results will be seen in the decays
of $\Upsilon\to \gamma f_J$ and $J/\psi\to \gamma f_J$. As the cases
of $\chi_{bJ}$ decays, especially for the process $\chi_{bJ}\to
\rho(\omega,\phi)\gamma$, QED process may give dominant
contributions.

\begin{acknowledgments}
We would like to thank Ce Meng for valuable discussions. This work
was supported in part by the National Natural Science Foundation of
China (No 10421503, No 10675003), the Key Grant Project of Chinese
Ministry of Education (No 305001), and the Research Found for
Doctorial Program of Higher Education of China.
\end{acknowledgments}


\end{document}